# Positron Induced Electron Emission from Graphite

A. J. Fairchild[1,a)], V. A. Chirayath[1], R. W. Gladen[1], A. R. Koymen[1] and A. H. Weiss[1]

[1]*University of Texas at Arlington, Department of Physics, 502 Yates St., Arlington, TX 76019.*

a)Corresponding author: alexander.fairchild@mavs.uta.edu

**Abstract.** In this paper, we present and analyze measurements of the positron induced electron spectra (PIES) from highly oriented pyrolytic graphite (HOPG). The spectra were obtained using a time of flight spectrometer attached to a variable energy positron beam. In the first measurements presented, the system was configured to obtain high resolution data from the annihilation induced KVV Auger transition of carbon. In the second set of data presented, PIES spectra were obtained for 3 different positron beam energies (1.25 eV, 3.5 eV and 4.5 eV). The resulting time of flight (ToF)-PIES exhibit contributions arising from either positron annihilation induced Auger processes (PAES), Auger mediated positron sticking (AMPS), or secondary electron emission. Our analysis indicates that for incident positron energies 3.5 eV and less, the ToF-PIES can be accounted for considering only two mechanisms: positron annihilation induced Auger processes or positron sticking.

## INTRODUCTION

Low energy positron bombardment can liberate electrons from a solid through three major mechanisms: a positron annihilation induced Auger process, Auger mediated positron sticking (AMPS) or secondary electron emission [1]. In positron annihilation induced Auger electron emission, surface trapped positrons annihilate with a surface bound electron leaving an electron hole. This sudden removal of an electron leaves the material in an excited state. One possibility for the system to reduce its energy is to emit an electron via an Auger transition. This Auger transition occurs when a less tightly bound electron comes to occupy the energy level of the hole, coupling the energy associated with this filling of the hole to another electron, which escapes into the vacuum. Recently, our group was the first to directly observe such an Auger transition occurring entirely within the valence band of graphene [2]. Our systems capability of efficiently transporting extremely low energy positrons (∼1 eV) to the sample made possible the direct observation of valance band Auger transitions by completely removing the obscuring presence of secondary electrons typically seen in similar measurements using photons or electrons.

In Auger mediated positron sticking (AMPS), the positron makes a transition from a positive energy scattering state to an image potential induced surface bound state. The energy associated with this transition is coupled to a valence electron inside the material which can then escape to the vacuum, provided it has received sufficient energy, through the AMPS process, to overcome the electron work function of the material. AMPS has allowed direct investigation of positron trapping at the surfaces of Cu, Au, and Bi2Te2Se [1, 3] and has motivated renewed interest in developing first principle theoretical models of positron surface states [4]. The maximum kinetic energy, $E_{max}^{AMPS}$, of an electron emitted as a result of AMPS is given by:

$$E_{max}{}^{AMPS} = E_p + E_{ss} - \varphi^- \qquad (1)$$

where $E_p$ is the incident positron energy, $E_{ss}$ is the positron surface state binding energy, and $\varphi^-$ is the electron work function. Therefore, a minimum incident positron energy of $\varphi^- - E_{ss}$ is required for the AMPS process to eject an electron.

AMPS can be distinguished from a secondary electron emission process in which the final state of the positron is a bulk state not a surface state. The maximum kinetic energy, $E_{max}^{SE}$ for this secondary electron emission process is given by:

$$E_{max}{}^{SE} = E_p + \varphi^+ - \varphi^- \qquad (2)$$

where $\varphi^+$ is the positron work function. Therefore, the minimum incident positron energy required for secondary electron emission is $\varphi^- - \varphi^+$. For highly oriented pyrolytic graphite (HOPG), previous measured values of $\varphi^+$ and $\varphi^-$ are 1.5 eV and 4.7 eV respectively [5]. Assuming a value of $E_{ss}$ = 2.4 eV, the AMPS and secondary electron threshold values are 2.3 eV and 3.2 eV respectively. Equations 1 and 2 also indicate that for incident positron beam energies of 1.25 eV, only Auger related processes supply enough energy to free electrons from a solid. In this paper, we present and analyze measurements of the time of flight positron induced electron emission spectra (ToF-PIES) from the surface of HOPG for three incident positron beam energies 1.25 eV, 3.5 eV and 4.5 eV. The 3.5 eV and 4.5 eV experimental spectra are analyzed in terms of two modelled spectra: an annihilation induced Auger electron spectra and an AMPS and secondary electron emission spectra.

## EXPERIMENTAL SETUP

The experimental apparatus has been described in detail elsewhere [6] so only a brief description will be given here. The positron beam system at the University of Texas at Arlington (UTA) consists of magnetically guided positrons from a Na22 source moderated using a thin sheet of tungsten, coupled to a time of flight (ToF) spectrometer consisting of a set of ExB plates, a relatively field free ToF tube, a micro-channel plate (MCP) detector, a BaF2 and a NaI scintillator (Fig. 1). The ToF of the emitted electrons is determined from the timing difference between the detection of the 511 keV annihilation gamma ray, by the BaF2 detector, and the detection of the electron by the MCP 1 meter away. The total number of positrons arriving at the sample is measured using the NaI detector and used to normalize the ToF spectra. The incident energy of the positrons was varied by applying a negative voltage bias to the sample. Lastly, the ToF tube was biased positively to act as a retarding field analyzer, to measure the incident positron beam energy, or biased negatively, to increase the energy resolution of the system or to remove unwanted portions of the electron ToF spectrum.

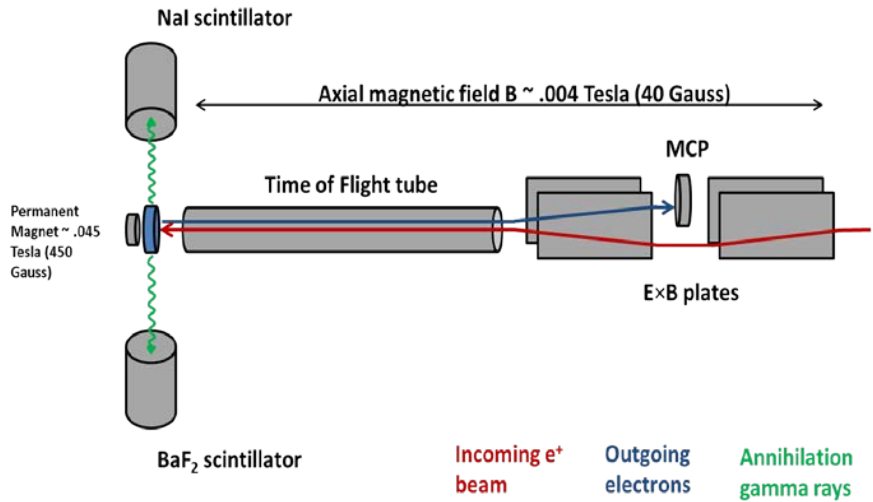

**FIGURE 1.** Schematic of the time of flight positron beam system at the University of Texas at Arlington. The red lines represent the incoming positron beam, the blue lines the outgoing electrons, and the green lines the 511 keV positron-electron annihilation gamma rays detected by the scintillators.

## RESULTS AND DISCUSSION

The time of flight positron induced electron spectra (ToF-PIES) for highly oriented pyrolytic graphite (HOPG) are presented in Figs. 2 and 3. The left panel of figure 2 is a high resolution measurement made with an incident positron beam energy of 17 eV, a sample bias of -30 V, a ToF tube bias of -180 V, and a magnetic field of ∼.18 Tesla at the sample surface. The ToF bias prevents any electrons below 180 eV from reaching the micro-channel plate (MCP) and slows down any higher energy electrons so that they spend more time in the field free region of the ToF tube thereby increasing their timing separation and improving the spectrometers energy resolution. The spectrum shows peaks corresponding to Auger transitions initiated by annihilation induced holes in the 1s core levels of carbon and oxygen (C KVV at ∼263 eV and O KVV at ∼503 eV). The third peak left unidentified in the spectrum is presently under investigation, but is believed to be due to positrons generating secondary electrons at the entrance of the ToF tube. We note that the integrated C KVV Auger intensities of figure 2 and panel (a) of figure 3 are within statistical error (2.09E-4 and 2.15E-4 counts per positron respectively). This is consistent with the hypothesis that at higher positron implantation energies the implanted positrons diffuse back and become trapped at the surface before annihilation. The right panel of figure 2 is a comparison between electron-excited Auger electron spectroscopy (EAES) [7] and the energy converted positron annihilation induced Auger electron spectroscopy (PAES) carbon KVV Auger line shape of the left panel. The EAES line shape been scaled to agree at the peak of the PAES line shape and shifted to lower energy by 13 eV to aid in comparison of the line shapes. Good overall agreement between our results and previous high resolution EAES measurements of the C KVV Auger line shape demonstrates that our ToF spectrometer is capable of similarly high resolution measurements. Furthermore, the agreement between PAES and EAES supports the hypothesis that the Auger process is similar in the positron and electron case.

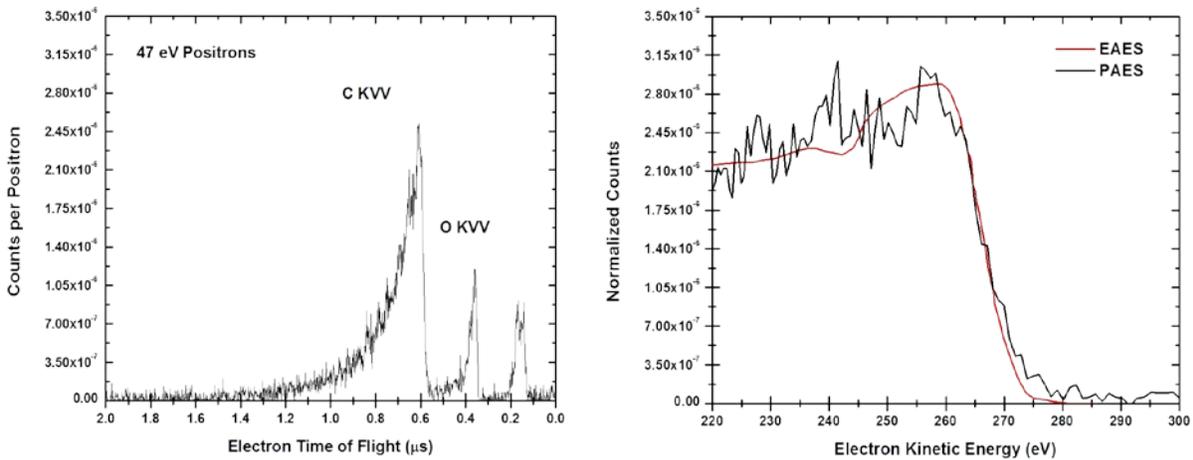

**FIGURE 2.** High resolution time of flight positron induced electron spectra (ToF-PIES) from HOPG. The left panel is the ToF-PIES spectrum for 47 eV incident positrons using ToF positron beam settings that give increased energy resolution at the C KVV Auger peak (∼263 eV). Sharp, well separated peaks corresponding to KVV Auger transitions in carbon and oxygen are visible. The right panel is a comparison between the carbon KVV line shape obtained by electron-excited Auger electron spectroscopy (EAES) [7] and by positron annihilation induced Auger electron spectroscopy (PAES).

Figure 3 is ToF-PIES measurements of HOPG made with the ToF tube grounded, to permit the transport of all electrons to the MCP. The positron beam energy distribution, measured at the ToF tube, was ∼1 eV. Incident positron beam energies of 1.25 eV, 3.5 eV and 4.5 eV were obtained by biasing the sample to -0.25 V, -2.5 V and -3.5 V respectively. Panel (a) shows the 0.25 V sample bias (1.25 eV positron energy) experiment where core (C KVV and O KVV) and valence (VVV) Auger processes are the only energetically permitted electron emission mechanisms. In order to obtain the effect the negative sample bias has on the valence and core Auger peaks, the 0.25 V spectrum was first scaled by the ratio of the integrated C KVV intensity of either the 2.5 V or 3.5 V spectra, to the integrated C KVV intensity of the 0.25 V spectrum. This sample bias dependent C KVV area normalization was done because the fraction of incident positrons that form positronium, which changes considerably as a function of the incident positron energy, competes with the number of positrons that trap and annihilate at the surface [8]. Next, using an experimentally derived electron ToF to kinetic energy conversion function, the scaled spectra were converted into energy, shifted in energy by the difference in sample biases (2.25 V or 3.25 V), then converted back into electron ToF. The advantage of viewing the spectra in ToF is that the structure of the low energy VVV peaks may be seen more clearly than in the energy spectra. Another advantage is that Auger peaks separated by hundreds of eV may be displayed conveniently on one scale. The overall effect of the sample bias is negligible for the core Auger peaks (C KVV and O KVV) but the valence Auger peak (VVV) is narrowed considerably.

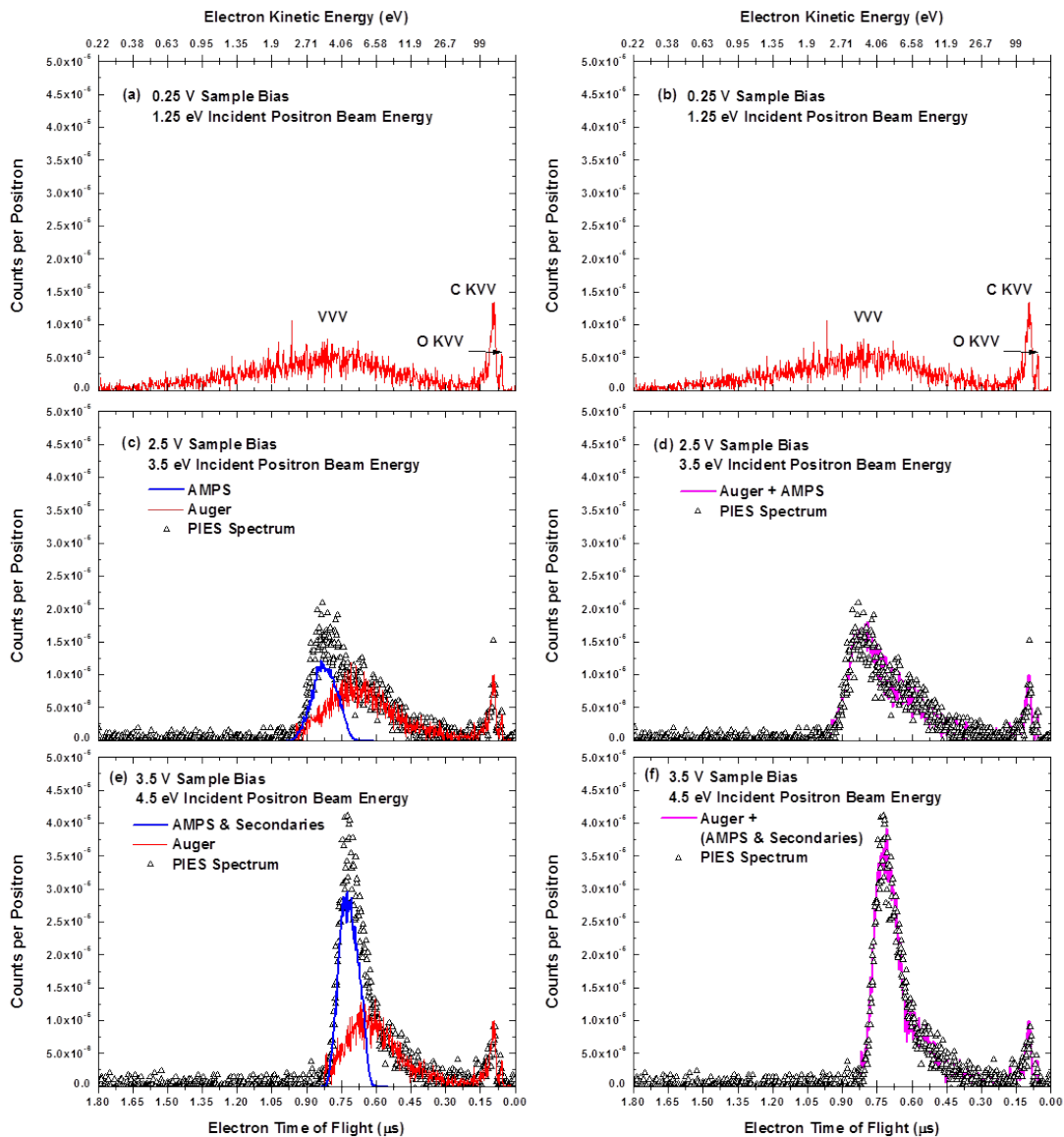

**FIGURE 3.** Time of flight positron induced electron spectra (ToF-PIES) from HOPG for several incident positron energies. Note that the ToF axis has been reversed so that the ToF scale goes from higher ToF (lower energy) to lower ToF (higher energy). Panel (a) and panel (b) are the ToF-PIES spectrum for 0.25 V sample bias (1.25 eV positrons) where only core (C KVV and O KVV) and valence (VVV) Auger processes are energetically possible. Panels (c) and (e) are comparisons between the sample bias shifted Auger spectrum from panel (a) in red, an instrumentally broadened theoretical Auger mediated positron sticking (AMPS) spectrum in blue and the experimental spectrum in black open triangles. Panels (d) and (f) are comparisons of the experimental spectrum with the sum of the Auger and AMPS or Auger and (AMPS and Secondaries) spectra in magenta.

Figure 3 panels (c) and (e) are ToF-PAES measurements of HOPG for 2.5 V and 3.5 V sample biases (3.5 eV and 4.5 eV incident positron energies) respectively in black open triangles shown alongside an instrumentally broadened theoretical AMPS spectra in blue and the corresponding sample bias shifted valence and core Auger spectrum in red.

The theoretical AMPS spectra is based on a model which incorporates first principle calculations of the electronic density of states near the Fermi level, together with estimates of the probabilities of electron escape and positron sticking and the measured positron energy distribution. The details and results of the AMPS line shape modelling will be presented in a forthcoming publication [9]. The theoretical AMPS spectra has been broadened to account for the effect of the spectrometer response function following previous work [2, 10]. For panel (e), the AMPS model was adjusted to account for electrons emitted via secondary electron processes in which the final state of the positron is the bulk state not the surface state. A positron final state energy was used in the model that was a mixture of the bulk state and surface state positron energies. The experimental line shape is expected to become less AMPS like and more secondary like as the incident positron beam energy is increased due to the decreasing probability for positron sticking [1].

Figure 3 panels (d) and (f) show the simple sum (magenta) of the instrumentally broadened theoretical spectra (blue) and the corresponding Auger spectra (red) alongside the ToF-PAES spectra (open black triangles) for 2.5 V and 3.5 V sample biases (3.5 eV and 4.5 eV incident positron energies). The excellent agreement demonstrates the success of the model in reproducing the experimental line shape and confirms our hypothesis, that at incident positron energies 3.5 eV and below the only channels available for electron emission are annihilation induced Auger processes and AMPS. Finally, since only a simple scaling by the ratio of the integrated C KVV Auger peak intensities, in order to correct for the increasing positronium formation at higher sample biases, was required to match the higher energy side of the low energy peak. And since annihilation induced Auger processes directly compete with positronium formation, the agreement provides additional evidence that the broad low energy peak in the 0.25 V ToF-PAES data is indeed related to an annihilation induced valence Auger process (VVV).

## SUMMARY

We have presented measurements of the kinetic energy distribution of electrons emitted from the surface of highly oriented pyrolytic graphite (HOPG) as a result of positron bombardment using the University of Texas at Arlington's (UTA) time of flight (ToF) positron beam system. The time of flight positron induced electron spectra (ToF-PIES) presented contain Auger peaks initiated by valence band and K-shell annihilations, as well as peaks due to Auger mediated positron sticking (AMPS). An analysis of the TOF-PIES spectra taken at 3 different positron beam energies suggests that by controlling the energy of the incident positron beam the relative contributions to the ToF-PIES from Auger processes, AMPS and secondary electron emission can be studied separately using one experimental setup. The 3.5 eV and 4.5 eV experimental spectra were decomposed into two modelled spectra, one containing electrons emitted as a result of positron annihilation induced Auger processes and another containing electrons emitted as a result of AMPS and secondary electron emission processes. The agreement between the experimental spectra and the modelled spectra provides strong confirmation that for incident positron beam energies 3.5 eV and below, the only two mechanisms for electron emission are PAES and AMPS. And, for an incident positron beam energy of 1.25 eV only Auger related processes are possible.

## ACKNOWLEDGMENTS

This work was supported by NSF grants DMR 1508719 and DMR 1338130 and Welch Foundation grant No. Y-1968-20180324.

## REFERENCES

[1]   S. Mukherjee et al., Phys. Rev. Lett. **104**, p. 247403 (2010).
[2]   V. A. Chirayath et al., Nat. Commun. **8**, p. 16116 (2017).
[3]   V. Callewaert et al., Phys. Rev. B **94**, p. 115411 (2016).
[4]   V. Callewaert et al., Phys. Rev. B **96**, p. 085135 (2017).
[5]   P. Sferlazzo et al., Phys. Rev. Lett. **60**, 538–541 (1988).
[6]   S. Mukherjee et al., Rev. Sci. Instrum. **87**, p. 035114 (2016).
[7]   J. E. Houston et al., Phys. Rev. B **34**, 1215–1226 (1986).
[8]   V. A. Chiriyath et al., AIP Conf. Proc. **This Volume** (Positronium formation in graphene and graphite).
[9]   A. J. Fairchild et al., Manuscript in preparation .
[10]  A. J. Fairchild et al., J. Pys.: Conf. Ser. **791**, p. 012030 (2017).